\documentclass[9pt,conference]{IEEEtran}

\usepackage{amsmath,amssymb,amsfonts,amsthm,enumitem} 

\usepackage{amssymb}
\usepackage{graphicx}
\usepackage{textcomp}
\usepackage{makecell}
\usepackage{float}
\usepackage{placeins}
\usepackage{multirow}
\usepackage{dblfloatfix}
\usepackage{pifont}
\usepackage{colortbl}
\usepackage[normalem]{ulem}
\useunder{\uline}{\ul}{}
\usepackage{booktabs}
\usepackage{balance}
\usepackage{comment}
\usepackage{subcaption}
\usepackage{hyphenat}
\usepackage{amsmath}
\usepackage[monochrome]{xcolor}

\usepackage{algorithm}
\usepackage{algpseudocode}
\usepackage{siunitx}
\usepackage{url}

\algrenewcommand\algorithmiccomment[1]{\texttt{// #1}}

\AtBeginDocument{%
  \providecommand\BibTeX{{%
    \normalfont B\kern-0.5em{\scshape i\kern-0.25em b}\kern-0.8em\TeX}}}

\begin{document}



\title{FlipLLM: Efficient Bit-Flip Attacks on \textcolor{blue}{Multimodal LLMs} using Reinforcement Learning}

\author{\IEEEauthorblockN{Khurram Khalil, Khaza Anuarul Hoque}
\IEEEauthorblockA{\textit{Department of Electrical Engineering and Computer Science}\\ \textit{University of Missouri-Columbia, USA}\\
\{khurram.khalil, mkfqm, hoquek\}@missouri.edu}
}  
\maketitle

\begin{abstract}
Generative Artificial Intelligence Models like Large Language Models (LLMs) and Large Vision Models (VLMs) exhibit state-of-the-art performance across a wide range of tasks but remain vulnerable to hardware-based threats, specifically bit-flip attacks (BFAs), posing a serious risk to their security in safety-critical applications. Existing BFA discovery methods—gradient-based, static analysis, and search-based—lack generalizability and struggle to scale, often failing to analyze the vast parameter space and complex interdependencies of modern foundation models in a reasonable time. This paper proposes \emph{FlipLLM}, a reinforcement learning (RL) architecture-agnostic framework that formulates BFA discovery as a sequential decision-making problem. FlipLLM combines sensitivity-guided layer pruning with Q-learning to efficiently identify minimal, high-impact bit sets capable of inducing catastrophic failure. We demonstrate the effectiveness and generalizability of FlipLLM by applying it to a diverse set of models, including prominent text-only LLMs (GPT-2 Large, LLaMA 3.1 8B, and DeepSeek-V2 7B), VLMs such as LLaVA 1.6, and datasets, such as MMLU, MMLU-Pro, VQAv2, and TextVQA. Our results show that FlipLLM can identify critical bits that are vulnerable to BFAs up to \textbf{2.5$\times$} faster than SOTA methods. We demonstrate that flipping the FlipLLM-identified bits plummets the accuracy of LLaMA 3.1 8B from 69.9\% to $\approx$ 0.2\%, and for LLaVA's VQA score from 78\% to almost 0\%, by flipping as few as 5 and 7 bits, respectively. 
\textcolor{blue}{Further analysis shows that applying standard hardware protection mechanisms, such as ECC SECDED, to the FlipLLM-identified bit locations completely mitigates the BFA impact, demonstrating the practical value of our framework for guiding hardware-level defenses.}
FlipLLM offers the first scalable and adaptive methodology for exploring the BFA vulnerability of both language and multimodal foundation models, paving the way for comprehensive hardware-security evaluation.
\end{abstract}


\section{Introduction}
\label{sec:introduction}

Generative Artificial Intelligence models like Large Language Models (LLMs)~\cite{brown2020language}  and Large Vision Models (VLMs)  represent a transformative advancement in artificial intelligence, finding integration into mission-critical systems spanning healthcare, finance, and autonomous navigation~\cite{radford2019language, bommasani2021opportunities}. Their effective deployment mandates reliable and secure operation across diverse hardware infrastructures, from expansive cloud accelerators to resource-constrained edge devices. Unfortunately, recent works, such as ~\cite{das2024attentionbreaker, sharma2025prisonbreak} have shown that LLMs are vulnerable to \emph{targeted}  bit-flip attacks (BFAs), where manipulating a minimal and carefully selected subset of parameter bits—a negligible fraction of the total—can lead to a catastrophic functional collapse, rendering the model unusable for its intended task.

Existing techniques for identifying critical bits susceptible to BFAs encounter substantial difficulties when scaled to modern foundation models. Gradient-based methods, exemplified by DeepHammer~\cite{yao2020deephammer}, leverage parameter gradients but often struggle with the inherently discrete nature of bit alterations and may fail to capture the intricate, non-linear parameter interactions characteristic of transformer architectures. Alternatively, direct search-based approaches, encompassing progressive exploration~\cite{wang2020progressive} or fault sneaking heuristics~\cite{zhao2019fault}, face insurmountable computational costs and protracted execution times due to the combinatorial explosion of possible bit locations within billion-parameter models. Recently, researchers applied BFAs to state-of-the-art LLMs. For instance, authors in \cite{sharma2025prisonbreak} employed static analysis for identifying the critical bits. In contrast, the authors in~\cite{das2024attentionbreaker} have utilized evolutionary algorithms. Both of these works offer valuable contributions but pose significant limitations. Static analysis inherently lacks adaptability to diverse or novel architectures (e.g., Mixture of Experts \cite{deepseek2024}). On the other hand, evolutionary strategies depend on predefined genetic operators, which may not be optimally suited for the unique optimization landscape of every model and can still incur significant computational overhead. Hence, a significant research gap exists for a practical BFA methodology that offers both high efficiency and inherent adaptability to the diverse and rapidly evolving foundation model ecosystem.

\emph{Contributions:} To address these challenges, we introduce \textbf{FlipLLM}, a reinforcement learning (RL) based framework that reformulates the BFA optimization challenge as a sequential decision-making problem. FlipLLM combines sensitivity-guided search space pruning with Q-learning to learn efficient policies that identify sparse sets of critical bits responsible for functional collapse. Unlike prior techniques~\cite{das2024attentionbreaker, sharma2025prisonbreak}, FlipLLM is both architecture-agnostic and computationally efficient, enabling rapid evaluation of robustness and fault resilience without exhaustive search or heuristic tuning. Specifically, FlipLLM adaptively learns vulnerable bit combinations with minimal perturbation by relying on a hybrid sensitivity analysis mechanism that integrates static (weight magnitude) and dynamic (gradient) information to guide effective search space pruning while avoiding expensive enumeration, reliance on fixed genetic operators, and exhaustive search. We demonstrate the scalability and generalizability of FlipLLM by successfully attacking a diverse set of text-only LLMs (GPT-2 Large, LLaMA 3.1 8B, DeepSeek-V2 7B) and VLMs like LLaVA 1.6. Furthermore, we identify architectural fault localization patterns across these diverse models, revealing a consistent concentration of vulnerabilities in attention projections and normalization parameters, providing actionable insights for developing more efficient and targeted attack strategies. 
\textcolor{blue}{It is worth mentioning that this paper does not re-establish known BFA mechanisms such as RowHammer~\cite{lin2025gpuhammer}, but rather addresses a different problem: \emph{'how to efficiently identifying minimal, high-impact bit sets capable of inducing catastrophic failure in LLMs and VLMs?'} The proposed architecture-agnostic method enables hardware designers to: (i) assess model vulnerabilities by determining the number of critical bits in \textbf{minimal} time, and (ii) identify the precise locations of these bits to inform \textbf{cost-effective} protection strategies.}

\emph{Key Results:}
FlipLLM exposes the significant vulnerability of LLaMA 3.1 8B on the MMLU and MMLU-Pro benchmarks by identifying and flipping only 5 bits (a $6.23 \times 10^{-8}\%$ perturbation), achieving this \emph{2.5$\times$} faster than SOTA methods, such as GenBFA~\cite{das2024attentionbreaker} and collapsing accuracy from 69.9\% to $\approx$ 0.2\%. For the multimodal domain, FlipLLM collapses the VQA score of LLaVA 1.6 on the VQAv2 and TextVQA datasets from over 78\% to near 0\% with a similarly small set of 7 bit-flips. Furthermore, analysis of the learned policies reveals a consistent localization of vulnerable bits within architectural components, such as attention projections and layer normalization, providing new insights into the structural vulnerability patterns of modern foundation models. 
\textcolor{blue}{Further analysis shows that applying hardware protection mechanisms such as, ECC SECDED to the FlipLLM-identified critical bits completely mitigates BFA impact, maintaining near-baseline accuracy (e.g., 69.8\% vs. 69.9\% compared to catastrophic failure 0.21\% on LLaMA 3.1 8B in unprotected systems), demonstrating the practical value of our framework for guiding hardware-level defenses.}

The remainder of this paper is organized as follows: Section~\ref{sec:background} reviews background and related work. Section~\ref{sec:methodology} presents the FlipLLM framework. Section~\ref{sec:evaluation} outlines the experimental setup, Section~\ref{sec:results} discusses the results, and Section~\ref{sec:conclusion} concludes the paper with future directions.

\begin{figure*}[!t]
    \centering
    \includegraphics[width=0.8\textwidth]{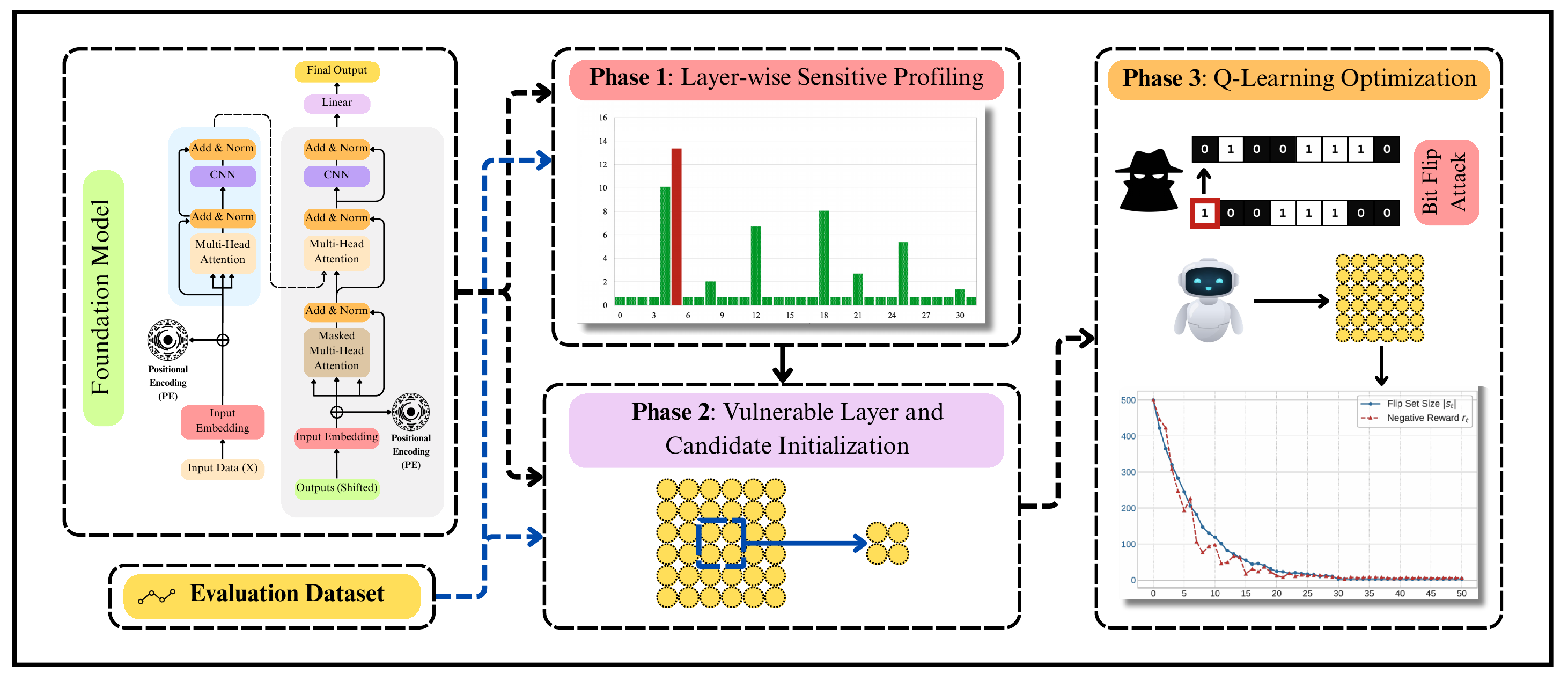}
    \caption{Overview of the FlipLLM framework.}
    \label{fig:framework}
\end{figure*}

\section{Background and Related Work}
\label{sec:background}
This section outlines the background of BFAs as a critical threat to neural networks and, by extension, large foundation models. It further surveys the state-of-the-art approaches for identifying BFA vulnerabilities. We then provide a critical analysis of the limitations inherent in these techniques—most notably their lack of scalability and adaptability—which motivates the development of our reinforcement learning–based framework, FlipLLM.


\subsection{Bit-Flip Attacks Against Neural Networks}

The use of hardware vulnerabilities to compromise machine learning models gained prominence with Rakin et al.~\cite{rakin2019bit}, who showed that flipping a small number of strategically chosen bits in neural network weights (e.g., in DNNs/CNNs) could catastrophically degrade accuracy. This finding spurred extensive research into BFAs, targeting diverse goals such as untargeted performance collapse, targeted misclassification, and more recently, advanced manipulations like model jailbreaking~\cite{sharma2025prisonbreak}.
Existing methodologies developed to identify these critical bits for BFAs generally fall into two main paradigms. Gradient-based approaches estimate bit sensitivity via loss gradients~\cite{yao2020deephammer, liu2021gradient}, but their effectiveness is often limited by the non-differentiable nature of discrete bit flips and the complex, non-linear error landscape of modern models. In contrast, search-based techniques directly probe the parameter space—ranging from heuristic~\cite{wang2020progressive, zhao2019fault} to evolutionary algorithms—but suffer from poor scalability as model size increases, often requiring prohibitive computational time to find impactful attack vectors.

\subsection{Challenges and Prior Art in Attacking Large Foundation Models}
The application of BFAs to contemporary foundation models, including both LLMs and VLMs, significantly amplifies the challenges faced by existing methodologies. The sheer scale of these models, often comprising billions of parameters, renders brute-force or exhaustive search techniques computationally infeasible. Furthermore, the intricate architectures, characterized by complex self-attention mechanisms, residual connections, and normalization layers, result in highly non-linear dependencies between parameters and overall model behavior. This complexity makes predicting the precise impact of bit-flips exceedingly difficult using simplified models or heuristics.

Recognizing these challenges, recent research has focused specifically on BFAs against large models. Notable examples include \textit{PrisonBreak}~\cite{sharma2025prisonbreak}, which employed static analysis techniques to identify potential vulnerabilities, and \textit{GenBFA}~\cite{das2024attentionbreaker}, which utilized evolutionary algorithms to search for effective bit-flips. While representing important progress, these approaches exhibit certain limitations. Static methods, by relying on pre-computed parameter importance, lack adaptability to dynamic, input-dependent vulnerabilities and struggle to generalize across diverse architectures (e.g., text-only vs. multimodal, or emerging designs like Mixture of Experts~\cite{deepseek2024}). Evolutionary methods, although more flexible, rely on heuristic-based genetic operators and typically incur high runtime due to the repeated evaluation of a large population of candidates, limiting their efficiency.
\textcolor{blue}{Recent work by Almalky et al.~\cite{almalky2025} investigated the applicability of vision-domain BFA techniques to LLMs, finding that conventional attacks like ``BFA" and ``Deep-TROJ" exhibit significantly reduced effectiveness compared to their performance on DNNs. Their systematic evaluation across multiple LLM architectures revealed that BFA often fails to degrade model performance even after hundreds of iterations, while trojan insertion methods struggle to maintain clean accuracy alongside backdoor functionality. These findings highlight fundamental differences in how LLMs respond to weight perturbation compared to vision models.}

Overall, existing techniques either struggle to generalize across architectures or require significant computational resources to discover minimal yet impactful bit flips. This reveals a clear gap: the need for an architecture-agnostic, adaptive, and efficient framework capable of identifying critical bit vulnerabilities in the rapidly evolving landscape of foundation models. Our work, FlipLLM, addresses this deficiency by leveraging the adaptive, policy-driven search capabilities of RL, as detailed in the subsequent section.

\section{Methodology: The FlipLLM Framework}
\label{sec:methodology}



The proposed \textit{FlipLLM} framework, shown in Figure~\ref{fig:framework}, provides a structured and scalable approach for identifying minimal and high-impact bit-flip vulnerabilities in LLMs. The process begins with Phase 1, the \textit{Sensitivity Profiling Phase}, where a hybrid metric—combining static weight magnitudes and dynamic gradient information—ranks parameter importance on a layer-by-layer basis efficiently. This yields a reduced candidate set of parameters most susceptible to targeted perturbations. Next, in Phase 2, the \textit{Vulnerable Layer and Candidate Initialization Phase}, leverages this analysis to select the single most vulnerable layer and uses its highest-sensitivity parameters as a high-quality starting point for focused optimization. Finally, Phase 3, the \textit{Q-Learning Optimization Phase}, casts BFA discovery as a sequential decision-making problem. Here, a reinforcement learning (RL) agent intelligently explores the constrained search space, refining the candidate bit set via reward-guided exploration to identify a minimal, critical subset of bits. The output of FlipLLM is a compact, high-impact set of vulnerable bit indices that induce worst-case degradation with minimal perturbation. The following subsections detail the threat model, formal notation, and each algorithmic stage of FlipLLM.

\newcommand{\proc}[1]{\textsc{#1}}

\begin{algorithm}[!t]
\caption{Phase 1: Layer-wise Sensitivity Profiling}
\label{alg:sensitivity_profiling}
\begin{algorithmic}[1]
\footnotesize
\Require LLM \( M \), Tokenizer \( T \), Total Num Layers \( L \), Eval Data \( D_{\text{eval}} \), Sensitivity Mix \( \alpha \), Selection Rate \( r \)
\Ensure Layer Results List \( \text{LayerSensResults} \) containing \( (l, \emph{acc}^l, I_{\text{subset}}^l) \) tuples

\State \( \text{LayerSensResults} \leftarrow [~] \)

\For{\( l = 1 \) to \( L \)}
    \State \( W^l \leftarrow \proc{GetLayerWeights}(M, l) \)
    \State \( \nabla W^l \leftarrow \proc{CalculateGradients}(M, T, l, D_{\text{eval}}) \)
    \State \( W^l_N, \nabla W^l_N \leftarrow \proc{L2Normalize}(W^l, \nabla W^l) \)
    \Statex \hspace*{2em} \texttt{// Hybrid sensitivity score}
    \State \( S^l \leftarrow \alpha |\nabla W^l_N| + (1 - \alpha) |W^l_N| \)
    \Statex \hspace*{2em} \texttt{// \( r \) is a fraction in (0, 1]}
    \State \( k \leftarrow \lfloor r \cdot |W^l| / 100 \rfloor \)
    
    \State \( I_{\text{subset}}^l \leftarrow \proc{GetTopKIndices}(S^l, k) \)
    \State \( W_{\text{pert}}^l \leftarrow \proc{FlipMSB}(W^l, I_{\text{subset}}^l) \)
    \State \( M_{\text{temp}} \leftarrow \proc{SetLayerWeights}(M, l, W_{\text{pert}}^l) \)
    \State \( \emph{acc}^l \leftarrow \proc{EvaluateAccuracy}(M_{\text{temp}}, T, D_{\text{eval}}) \)
    \Statex \hspace*{2em} \texttt{// Restore original weights}
    \State \( M \leftarrow \proc{SetLayerWeights}(M, l, W^l) \)
    \State Append \( (l, \emph{acc}^l, I_{\text{subset}}^l) \) to \( \text{LayerSensResults} \)
\EndFor

\Statex \texttt{// Find layer with min acc after perturbation}
\State Find \( (l^*, \emph{acc}_{\text{min}}, I_{\text{hybrid}}^{l^*}) \in \text{LayerSensResults} \) such that \( \emph{acc}_{\text{min}} = \min_{(l, \emph{acc}^l, I_{\text{subset}}^l) \in \text{LayerSensResults}} \emph{acc}^l \)

\State \Return \( \text{LayerSensResults}, l^*, I_{\text{hybrid}}^{l^*} \)

\end{algorithmic}
\end{algorithm}

\subsection{Threat Model}
\label{sec:threat_model}

In Machine Learning as a Service (MLaaS) settings, the use of shared computational platforms for deploying contemporary foundation models, including both LLMs and VLMs, gives rise to significant security vulnerabilities. The shared nature of hardware components, such as last-level caches and main memory, makes these models susceptible to potential BFAs. Adversaries can leverage side channels to manipulate bits within the memory cells containing the model's parameters~\cite{shuvo2023survey}, even without explicit permissions to user data. The associated risks can be categorized into three tiers, with an advanced-level attacker possessing sophisticated techniques to induce severe outcomes, such as complete model corruption or advanced manipulations like model jailbreaking~\cite{sharma2025prisonbreak}.

Consistent with recent state-of-the-art BFA research~\cite{das2024attentionbreaker, sharma2025prisonbreak}, this work concentrates on an adversary with elevated privileges who achieves unauthorized access to the memory locations holding the model's weights~\cite{liu2017fault}. The primary objective of such an attacker is to achieve the most adversarial effect while altering the minimum number of bits possible. By leveraging fault injection methods, such as the well-documented RowHammer attack~\cite{mutlu2019rowhammer}, the adversary can selectively flip critical bits within the model's weights. \textcolor{blue}{Recent demonstrations validate the practical feasibility of GPU Rowhammer attacks~\cite{lin2025gpuhammer}, achieving bit flips in GDDR6 memory on NVIDIA A6000 GPUs. This threat is realistic in multi-tenant cloud GPU environments with time-sliced~\cite{nvidia_gpu_operator} or spatially-shared~\cite{copik2024mignificient} workloads, where attackers with user-level CUDA execution privileges can co-locate with victim models in the same DRAM banks. Rowhammer attacks have been demonstrated across diverse memory technologies including LPDDR4~\cite{yaglikci2021understanding} and GDDR5~\cite{kwong2020rambleed}, and recent work shows attacks remain feasible even with modern defenses like Target Row Refresh~\cite{frigo2020trrespass, jattke2022blacksmith}.
Indeed such manipulation can result in a wide range of performance degradation, from subtle to severe, which in turn compromises the model's overall reliability. This vulnerability poses a particular threat in mission-critical applications, including those in healthcare, finance, and autonomous systems, where dependable operation is essential. While prior work demonstrates the fault injection mechanism, FlipLLM addresses the complementary targeting problem: identifying which specific bits to flip for maximum impact with minimal perturbations.}

\subsection{Problem Formalization}
\label{subsec:notation}

Let a foundation model $M$ be defined by its weight parameters $W$, organized across $L$ layers, where $W^l$ denotes the weights of layer $l$. This model's performance on a task is measured by an evaluation metric $\emph{acc}(M, D_{\text{eval}})$ on a dataset $D_{\text{eval}}$. The BFA problem is to find a minimal set of bit indices $I_{\text{critical}}$ within the set of all addressable bits $\mathcal{B}(W)$ that maximizes performance degradation. Formally, this is an optimization problem as follows.
\begin{align}
\label{eq:bfa_objective}
\min_{I \subseteq \mathcal{B}(W)} \quad & |I| \\
\label{eq:bfa_constraint}
\text{s.t.} \quad & \emph{acc}(M^{\text{flip}}_I, D_{\text{eval}}) \le \tau,
\end{align}

where $M^{\text{flip}}_I$ is the model with bits at indices $I$ flipped, and $\tau$ is a threshold representing catastrophic failure (e.g., near-random accuracy). Due to the discrete, high-dimensional nature of $\mathcal{B}(W)$, this problem is computationally intractable to solve exactly. FlipLLM is designed to find an effective approximate solution.

For clarity, we define key notations used in our algorithms. The hybrid sensitivity score for parameters in layer $l$ is $S^l$, a convex combination of L2-normalized weight magnitudes ($|W_N^l|$) and gradients ($|\nabla W_N^l|$) with mixing coefficient $\alpha$. A selection rate $r$ determines the initial candidate parameter set $I_{\text{subset}}^l$. For direct comparison with prior work~\cite{rakin2019bit, das2024attentionbreaker}, we focus on Most Significant Bit (MSB) flips, which represent a worst-case numerical perturbation. The most sensitive layer, $l^*$, is the one exhibiting the lowest accuracy after flipping the MSBs of its most sensitive parameters. Its initial index set, $I_{\text{hybrid}}^{l^*}$, becomes the starting state $s_0$ for the RL agent.

The RL phase is a Markov Decision Process (MDP) with states $s_t$ (the current index set), actions $a_t \in \{\text{add}, \text{remove}, \text{shift}\}$, and a reward function $r_t$. The agent learns an action-value function $Q(s, a)$ over $G$ episodes using learning rate $\alpha_{rl}$, discount factor $\gamma$, and exploration rate $\epsilon$. The final output is the optimized critical index set $I_{\text{critical}}$.

\begin{algorithm}[!t] 
\caption{Phase 3: Q-Learning Optimization}
\label{alg:q_learning}
\begin{algorithmic}[1] 
\footnotesize
\Require LLM \( M \), Tokenizer \( T \), Target Layer Index \( l^* \), Initial Index Set \( s_0 = I_{\text{hybrid}}^{l^*} \), Eval Data \( D_{\text{eval}} \), RL Hyperparameters \( G, \epsilon, \alpha_{rl}, \gamma \), Action Space \( A \)
\Ensure Optimized Critical Index Set \( I_{\text{critical}} \)

\State Initialize Q-table \( Q(s, a) \leftarrow 0 \) for all \( s, a \) encountered
\State \( s_{curr} \leftarrow s_0 \)

\For{\( t = 1 \) to \( G \)}
    \Statex \hspace*{2em}\texttt{// Select action using epsilon-greedy policy}
    \State \( a_t \leftarrow \proc{SelectActionEpsilonGreedy}(s_{curr}, Q, \epsilon, A) \)
    
    \Statex \hspace*{2em}\texttt{// Update index set based on selected action}
    \State \( s_{next} \leftarrow \proc{TransitionState}(s_{curr}, a_t, A) \)

    \Statex \hspace*{2em}\texttt{// Get original layer weights}
    \State \( W_{\text{orig}}^{l^*} \leftarrow \proc{GetLayerWeights}(M, l^*) \)
    
    \Statex \hspace*{2em}\texttt{// Perturb layer weights based on selected indices}
    \State \( W_{\text{pert}}^{l^*} \leftarrow \proc{FlipMSB}(W_{\text{orig}}^{l^*}, s_{next}) \)

    \Statex \hspace*{2em}\texttt{// Set perturbed weights in model}
    \State \( M_{\text{temp}} \leftarrow \proc{SetLayerWeights}(M, l^*, W_{\text{pert}}^{l^*}) \)

    \Statex \hspace*{2em}\texttt{// Evaluate model accuracy with perturbed weights}
    \State \( \emph{acc}_t \leftarrow \proc{EvaluateAccuracy}(M_{\text{temp}}, T, D_{\text{eval}}) \)

    \Statex \hspace*{2em}\texttt{// Restore original weights}
    \State \( M \leftarrow \proc{SetLayerWeights}(M, l^*, W_{\text{orig}}^{l^*}) \)

    \Statex \hspace*{2em}\texttt{// Compute reward}
    \State \( r_t \leftarrow -(1 - \emph{acc}_t) / \max(1, |s_{next}|) \)
    
    \Statex \hspace*{2em}\texttt{// Q-update}
    \State \( Q(s_{curr}, a_t) \leftarrow (1 - \alpha_{rl}) Q(s_{curr}, a_t) + \alpha_{rl} (r_t + \gamma \max_{a'} Q(s_{next}, a')) \)

    \State \( s_{curr} \leftarrow s_{next} \)
\EndFor

\Statex \texttt{// Extract  optimal index set from final Q-table}
\State \( I_{\text{critical}} \leftarrow \proc{ExtractOptimalIndices}(s_{curr}, Q) \)

\State \Return \( I_{\text{critical}} \)

\end{algorithmic}
\end{algorithm}

\subsection{Phase 1: Layer-wise Sensitivity Profiling}
\label{subsec:sensitivity}
The initial phase of FlipLLM addresses the challenge of LLM parameter scale by efficiently identifying which layers of the LLM exhibit the highest sensitivity to bit-flip perturbations. This sensitivity profiling, detailed in Algorithm~\ref{alg:sensitivity_profiling}, constrains the search space for the subsequent, more computationally intensive optimization phase. To quantify sensitivity, we compute a \textit{FlipLLM Sensitivity Score}, \( S^l \), for every parameter within each layer \( l \). This metric is designed to capture a parameter's static importance, often correlated with its magnitude, and also its dynamic influence on the model's behavior, as reflected by its gradient during backpropagation. 
\textcolor{blue}{Sensitivity-based bit selection has been explored in prior BFA research through various approaches: gradient-based ranking in DeepHammer~\cite{yao2020deephammer} and the original BFA work~\cite{rakin2019bit}, magnitude-based heuristics in weight pruning literature, and more recently, hybrid scoring combining both gradient and magnitude information in GenBFA~\cite{das2024attentionbreaker}. While we adopt a similar hybrid formulation for initializing our search space, FlipLLM's contribution lies in reformulating BFA discovery as a sequential decision-making problem solved through RL, enabling adaptive exploration that learns synergistic bit combinations rather than relying on static sensitivity rankings or evolutionary operators.}
The score provides a unified measure of potential impact upon perturbation. Mathematically, the sensitivity score \( S^l \) is defined as a weighted combination:
\begin{equation}
    S^l = \alpha \cdot |\nabla W_N^l| + (1 - \alpha) \cdot |W_N^l|,
    \label{eq:sensitivity}
\end{equation}
In this equation, \( W_N^l \) and \( \nabla W_N^l \) are the L2-normalized vectors of weights and their corresponding gradients for layer \( l \), respectively (Algorithm~\ref{alg:sensitivity_profiling}, line 5). Normalization ensures fair comparison across parameters, potentially having vastly different scales. The coefficient \( \alpha \) balances the relative contributions of gradient versus magnitude information; its value is a configurable hyperparameter of the framework. Gradients are typically computed using a small but representative batch of data \( D_{\text{eval}} \) (line 4).

The profiling procedure iterates through each layer of the model (line 2). Within each iteration, the layer's weights \( W^l \) are extracted, and gradients \( \nabla W^l \) are computed (lines 3-4). Following normalization and calculation of sensitivity scores \( S^l \) (lines 5-6), a preliminary impact assessment is performed. A small fraction \( r \) of parameters exhibiting the highest sensitivity scores are identified, resulting in \( k \) indices \( I_{\text{subset}}^l \) (lines 7-8). To simulate hardware faults, the MSBs of these selected parameters are temporarily flipped (line 9). The functional consequence is measured by evaluating the accuracy \( \emph{acc}^l \) of the modified model (\(M_{\text{temp}}\)) on \( D_{\text{eval}} \) (lines 10-11). Crucially, original weights are restored (line 12) to ensure independent layer assessments. The accuracy \( \emph{acc}^l \) and indices \( I_{\text{subset}}^l \) are stored (line 13). Algorithm~\ref{alg:sensitivity_profiling} returns the results for all layers and identifies the most sensitive layer index \( l^* \) and its corresponding initial index set \( I_{\text{hybrid}}^{l^*} \) (line 15).

\begin{algorithm}[!b] 
\caption{FlipLLM Main Procedure}
\label{alg:FlipLLM_main}
\begin{algorithmic}[1] 
\footnotesize
\Require LLM \( M \), Tokenizer \( T \), Total Num Layers \( L \), Eval Data \( D_{\text{eval}} \), Sensitivity Mix \( \alpha \), Selection Rate \( r \), RL Hyperparameters \( G, \epsilon, \alpha_{rl}, \gamma \), Action Space \( A \)
\Ensure Final Critical Index Set \( I_{\text{critical}} \), Final Accuracy \( \emph{acc}_{\text{final}} \)

\Statex \Comment{\textbf{Phase 1:} Identify most sensitive layer}
\State \( \text{LayerSensResults}, l^*, I_{\text{hybrid}}^{l^*} \leftarrow \text{Algorithm~\ref{alg:sensitivity_profiling}}(M, T, L, D_{\text{eval}}, \alpha, r) \)

\Statex \Comment{\textbf{Phase 2:} Implicitly done by selecting \( I_{\text{hybrid}}^{l^*} \) as initial state for RL}
\State \( s_0 \leftarrow I_{\text{hybrid}}^{l^*} \)

\Statex \Comment{\textbf{Phase 3:} Optimize the index set using Q-Learning}
\State \( I_{\text{critical}} \leftarrow \text{Algorithm~\ref{alg:q_learning}}(M, T, l^*, s_0, D_{\text{eval}}, G, \epsilon, \alpha_{rl}, \gamma, A) \)

\State \( W_{\text{orig}}^{l^*} \leftarrow \proc{GetLayerWeights}(M, l^*) \)
\State \( W_{\text{final}}^{l^*} \leftarrow \proc{FlipMSB}(W_{\text{orig}}^{l^*}, I_{\text{critical}}) \)

\Statex \hspace*{2em}\texttt{// Set the perturbed weights in the model}
\State \( M_{\text{final}} \leftarrow \proc{SetLayerWeights}(M, l^*, W_{\text{final}}^{l^*}) \)

\State \( \emph{acc}_{\text{final}} \leftarrow \proc{EvaluateAccuracy}(M_{\text{final}}, T, D_{\text{eval}}) \)

\State \Return \( I_{\text{critical}}, \emph{acc}_{\text{final}} \)

\end{algorithmic}
\end{algorithm}


\subsection{Phase 2: Vulnerable Layer and Candidate Initialization}
\label{subsec:subset_selection}
This phase acts as a bridge between the broad, layer-level search and the fine-grained, bit-level optimization. Upon completion of Algorithm~\ref{alg:sensitivity_profiling}, the most sensitive layer $l^*$ and its initial sensitive index subset $I_{\text{hybrid}}^{l^*}$ are identified. Phase 2 consists of selecting this layer $l^*$ as the sole target for the BFA search and using its high-potential index set $I_{\text{hybrid}}^{l^*}$ as the starting state, $s_0$, for the RL agent as detailed in Algorithm~\ref{alg:q_learning}. This step is critical for scalability, as it leverages the global sensitivity analysis to focus the expensive RL exploration on the most promising region of the vast parameter space.

\subsection{Phase 3: Q-Learning Optimization}
\label{subsec:q_learning}
The central innovation of the FlipLLM framework lies in this third phase, which employs Q-learning to refine the initial candidate index set \( I_{\text{hybrid}}^{l^*} \) (provided as \(s_0\)) into a final, minimal set of highly impactful critical bit locations \( I_{\text{critical}} \). This phase adaptively explores complex interactions within layer \( l^* \). We formulate this as a MDP.

The \textit{state} \( s_t \) represents the current set of parameter indices designated for MSB perturbation, initialized as \( s_0 = I_{\text{hybrid}}^{l^*} \) (Algorithm~\ref{alg:q_learning}, line 2). The \textit{action} \( a_t \) is selected from \( A = \{\text{add}, \text{remove}, \text{shift}\} \) using an \(\epsilon\)-greedy strategy (line 4). 
\textcolor{blue}{This action space $A$ modifies the current bit index set as follows: \textit{add} selects a new parameter from the remaining sensitive set and adds its index to the current flip set; \textit{remove} eliminates a parameter index from the current flip set; \textit{shift} replaces one parameter index with another from the sensitive set while maintaining constant set size. The shift operation enables exploration of alternative parameter combinations without changing the cardinality of the flip set, allowing the RL agent to escape local optima where simply adding or removing indices proves insufficient. For example, if the current state $s_t = \{\text{idx}_1, \text{idx}_2, \text{idx}_3\}$ where each $\text{idx}_i$ represents the position of a parameter in the layer, and the sensitive set contains $\{\text{idx}_1, \ldots, \text{idx}_{100}\}$, a shift action might replace $\text{idx}_2$ with $\text{idx}_{47}$, yielding $s_{t+1} = \{\text{idx}_1, \text{idx}_{47}, \text{idx}_3\}$.}
This action transitions the state to \( s_{next} \) (line 5). The learning process is guided by the \textit{reward} function \( r_t \). After perturbing weights according to \( s_{next} \) (lines 6-8) and evaluating the resulting accuracy \( \emph{acc}_t \) (line 9), the reward is calculated (line 11):
\begin{equation}
\small
    r_t = -\frac{1 - \emph{acc}_t}{\max(1, |s_{next}|)} \label{eq:reward}
\end{equation}
This reward structure drives the agent towards discovering small (\(|s_{next}|\)) sets causing high impact (\(1-\emph{acc}_t\)). 
\textcolor{blue}{The reward function is intentionally designed for simplicity and directness. While more complex formulations could be explored, this straightforward design effectively captures the attacker's dual objective: maximizing model degradation (measured by $1 - \emph{acc}_t$) while minimizing the number of bit-flips. The negative reward structure naturally guides the RL agent toward sparse, high-impact fault combinations without requiring multi-objective optimization frameworks or manually tuned weighting parameters. Our empirical results (Section 5) demonstrate that this design achieves state-of-the-art performance, identifying critical fault sets with fewer bits than existing methods while maintaining 2.5× faster runtime, validating the effectiveness of this parsimonious formulation.}
The Q-learning algorithm updates the action-value function \( Q(s, a) \) using the Bellman equation (line 12):
\begin{equation}
\small
\textcolor{blue}{
\begin{aligned}
    Q(s_{\text{curr}}, a_t) \leftarrow (1 - \alpha_{rl}) Q(s_{\text{curr}}, a_t) + {} \\
    \alpha_{rl} \left( r_t + \gamma \max_{a'} Q(s_{\text{next}}, a') \right)
\end{aligned}
}
\label{eq:q_update}
\end{equation}
Original weights are restored after each evaluation (Algorithm~\ref{alg:q_learning}, line 10). The loop continues for \( G \) generations (line 3). Finally, the optimal set \( I_{\text{critical}} \) is extracted based on the learned Q-values (line 14).

The entire FlipLLM process is coordinated by the main procedure detailed in Algorithm~\ref{alg:FlipLLM_main}. This orchestrating algorithm first invokes the sensitivity profiling (Algorithm~\ref{alg:sensitivity_profiling}) to identify the most sensitive layer \( l^* \) and its initial high-potential index set \( I_{\text{hybrid}}^{l^*} \) (Algorithm~\ref{alg:FlipLLM_main}, line 2). This initial set \( I_{\text{hybrid}}^{l^*} \) is then used to initialize the starting state \( s_0 \) for the RL phase (line 5), effectively completing Phase 2. Subsequently, the Q-learning optimization (Algorithm~\ref{alg:q_learning}) is executed with \( s_0 \) and the identified layer \( l^* \) to yield the final, minimal set of critical indices \( I_{\text{critical}} \) (line 8). Finally, Algorithm~\ref{alg:FlipLLM_main} performs a concluding evaluation step: it applies the MSB flips precisely at the locations specified by the optimized set \( I_{\text{critical}} \) within the target layer \( l^* \) (lines 10-12) and measures the resulting final accuracy \( \emph{acc}_{\text{final}} \) (line 13). The procedure returns the identified critical index set \( I_{\text{critical}} \) and the final accuracy \( \emph{acc}_{\text{final}} \) achieved by perturbing only these critical bits (line 15).

\section{Evaluation Methodology}
\label{sec:evaluation}

This section details the empirical validation of the FlipLLM framework. We outline the experimental configurations, including target models, benchmarks, metrics, and implementation specifics, alongside the baseline methods used for comparative analysis, providing a reproducible foundation for the results presented in Section~\ref{sec:results}.

\subsection{Experimental Setup}
\label{subsec:eval_setup}

\subsubsection{Target Foundation Models}
We evaluate FlipLLM's effectiveness and generalizability on a diverse suite of pre-trained foundation models spanning both language-only and multimodal domains.
\begin{itemize}[leftmargin=*,itemsep=2pt]
    \item \textbf{Language Models (LLMs):} We target three LLMs offering diversity in architecture and parameter scale: GPT-2 Large (774M parameters)~\cite{radford2019language}, DeepSeek-V2 (a 7B Mixture-of-Experts model)~\cite{deepseek2024}, and LLaMA 3.1 8B Instruct (an 8B dense transformer model)~\cite{llama2024}.
    \item \textbf{Large Vision Models (VLMs):} To demonstrate FlipLLM's architecture-agnostic capabilities, we also attack LLaVA 1.6 (7B parameters)~\cite{LLaVA2024}, a leading vision-language model.
\end{itemize}
This selection allows for assessing FlipLLM's applicability across varied complexities, sizes, and modalities. All models were evaluated using 8-bit integer weight quantization (W8), reflecting common deployment scenarios that present opportunities for precise fault injection attacks. It is worth mentioning that we plan to release our code upon acceptance.

\subsubsection{Datasets and Evaluation Metrics}

To evaluate the functional impact of FlipLLM-induced bit-flips, we use comprehensive benchmarks tailored to each model's domain.
\begin{itemize}[leftmargin=*,itemsep=2pt]
    \item \textbf{For LLMs,} we use the Massive Multitask Language Understanding (MMLU) benchmark~\cite{hendrycks2021mmlu} and its more challenging MMLU-Pro variant~\cite{wang2024mmlupro}. These benchmarks measure knowledge and reasoning across 57 subjects with over 15,000 multiple-choice questions, and 14 subjects with 12,000 questions respectively, providing a robust measure of general cognitive capability. Performance is reported as the average 5-shot accuracy across all tasks.

    \item \textbf{For VLMs,} we evaluate vulnerability on two standard Visual Question Answering (VQA) benchmarks: VQAv2~\cite{vqav2_2017} and TextVQA~\cite{textvqa2019}. Performance is measured using the standard VQA score, which reflects accuracy on open-ended visual reasoning questions.
\end{itemize}

The primary evaluation metric for our attack is the final model performance (Accuracy or VQA Score) after applying the identified critical bit-flips, compared to the original model's baseline performance (\emph{Original Model Performance}). Critically, we measure the attack stealth and efficiency by the number of bit-flips required to achieve this degradation (\(|I_{\text{critical}}|\)). Though we use MMLU and MMLU-Pro for LLMS and VQAv2 and TextVQA for VLMS in this paper, our proposed method is generic and can be easily applied to any other evaluation dataset.

\subsection{FlipLLM Configuration and Implementation}
Experiments were conducted using PyTorch 2.1.0 with CUDA 12.1. The optimization and evaluation for all models were performed on a cluster node equipped with 4 NVIDIA A100 80GB GPUs. For the reported results, FlipLLM hyperparameters are fixed based on preliminary tuning: sensitivity mixing coefficient \(\alpha = 0.5\), subset selection rate \(r = 0.1\%\), Q-learning rate \(\alpha_{rl} = 0.1\), discount factor \(\gamma = 0.9\), exploration rate \(\epsilon = 0.1\), and \(G = 50\) training generations.
For sensitivity analysis and RL reward calculation, we use a randomly sampled subset of the evaluation data, \( D_{\text{eval}} \). For MMLU/MMLU-Pro, this consists of 1/10th of the subjects, with 100 questions per subject. For VQAv2/TextVQA, we use 1,000 randomly sampled questions from their respective validation splits.
The total computation time for the FlipLLM search across all four foundation models and their respective benchmarks, including experimentation and optimization, is approximately \textbf{850 GPU hours}.

\subsection{Comparison with Baseline and SOTA Methods}
FlipLLM's performance is evaluated against several baselines. First is the \emph{Original Model Performance} on the respective benchmarks, establishing the pre-attack capability. Second, we include a \emph{Random MSB Flips} baseline, where \(N_{rand} = 10,000\) bits are randomly flipped within the same layer \(l^*\) that FlipLLM identified as most sensitive. 
\textcolor{blue}{
This number was chosen based on two considerations: (1) it represents 2000$\times$ more flips than FlipLLM's targeted approach ($\approx$ 5 bits), providing strong statistical evidence that random injection fails despite overwhelming numerical advantage, and (2) computational constraints limit exhaustive random sampling in billion-parameter models. Even with 10,000 flips, the probability of randomly selecting FlipLLM's $\approx$ 5-bit critical set is approximately $\binom{6.4 \times 10^{10}}{5}^{-1} \approx 10^{-48}$, making random discovery effectively impossible.}
Third, we compare the computational time of FlipLLM against the state-of-the-art evolutionary BFA approach, \emph{GenBFA}~\cite{das2024attentionbreaker}. Additionally, we compare against a gradient-based baseline that implements the core methodology of DeepHammer~\cite{yao2020deephammer}, a seminal work in BFA. We exclude a direct comparison with \textit{PrisonBreak}~\cite{sharma2025prisonbreak} as its objective is model jailbreaking (generating unsafe content), which is a distinct attack goal from our focus on catastrophic accuracy collapse.


\begin{table*}[t]
    \centering
    \caption{FlipLLM Performance Summary: Minimal Bit Flips for Catastrophic Performance Collapse on Language and Vision-Language Benchmarks. Acc. = Accuracy, Crit. Bits = Average Number of Critical Flips Identified by FlipLLM.}
    \label{tab:main_results}
    \begin{tabular}{@{}llcccccc@{}}
        \toprule
        \multicolumn{2}{c}{\textbf{Model}} & & \multicolumn{2}{c}{\textbf{Primary Benchmark}} & \multicolumn{2}{c}{\textbf{Challenging Benchmark}} & \\
        \cmidrule(lr){4-5} \cmidrule(lr){6-7}
        \textbf{Type} & \textbf{Name} & \textbf{Parameters} & \textbf{Baseline Perf.} & \textbf{Final Perf.} & \textbf{Baseline Perf.} & \textbf{Final Perf.} & \textbf{Crit. Bits} \\
        \midrule
        \multirow{3}{*}{\textbf{LLM}} 
        & LLaMA 3.1 8B      & 8B         & 69.9\% (MMLU) & 0.21\% & 45.1\% (MMLU-Pro) & 0.15\% & \textbf{5} \\ 
        & DeepSeek-V2       & 7B (MoE)   & 71.3\% (MMLU) & 0.19\% & 48.2\% (MMLU-Pro) & 0.16\% & \textbf{6} \\ 
        & GPT-2 Large       & 774M       & 30.5\% (MMLU) & 0.35\% & 26.8\% (MMLU-Pro) & 0.25\% & \textbf{5} \\
        \midrule
        \multirow{1}{*}{\textbf{VLM}} 
        & LLaVA 1.6    & 7B    & 78.2\% (VQAv2) & 0.5\% & 62.5\% (TextVQA) & 0.3\% & \textbf{7} \\
        \bottomrule
    \end{tabular}
\end{table*}

\begin{figure}[!t]
    \centering
    \includegraphics[width=0.95\linewidth]{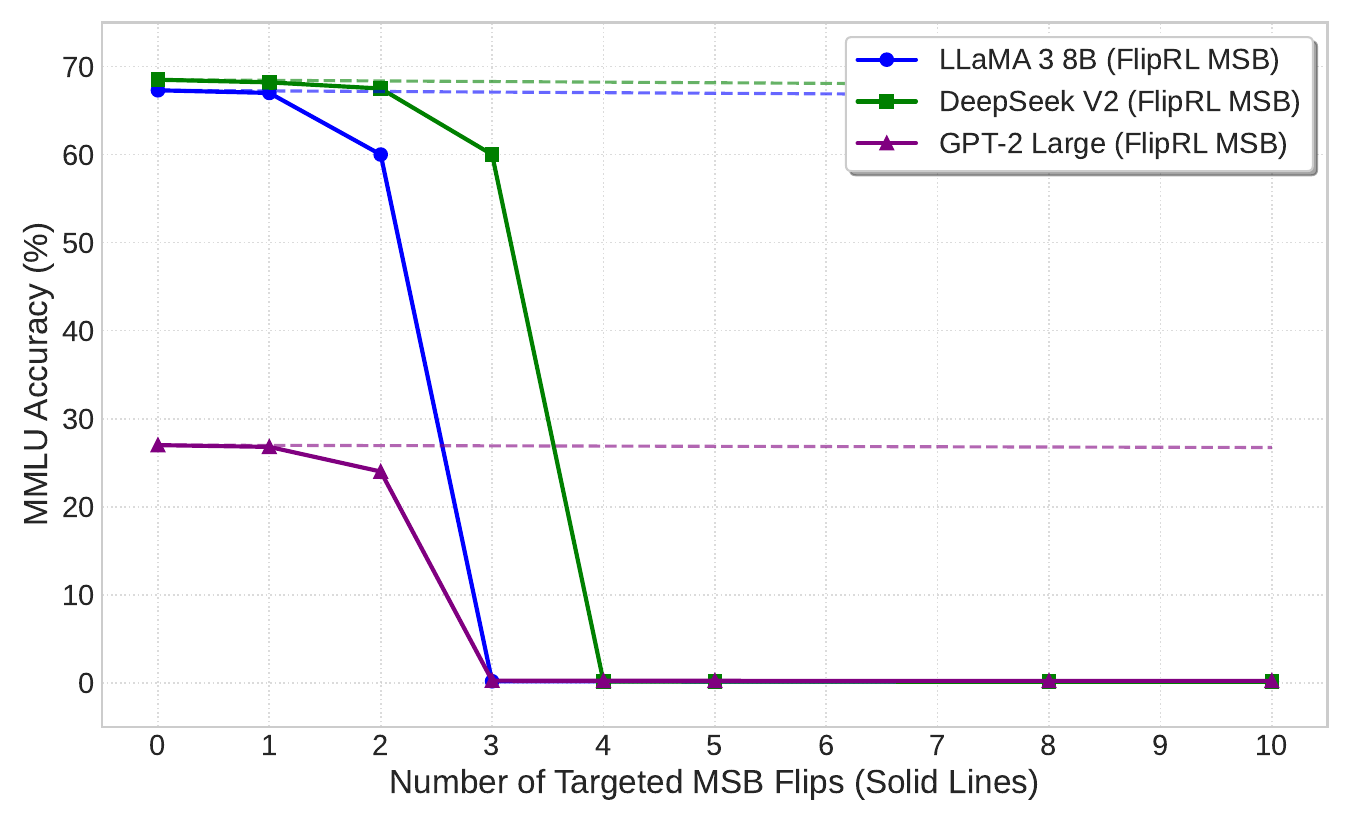}
    \caption{MMLU Accuracy vs. Number of bit flips for LLaMA 3.1 8B, DeepSeek V2 and GPT-2 Large model.
    }
    \label{fig:FlipLLM_vs_random_combined}
    \vspace{-2mm}
\end{figure}

\begin{table*}[t] 
    \centering
    \caption{Attack Efficiency Comparison of FlipLLM Against Baselines on LLM (MMLU Accuracy) and VLM (VQAv2 Score). Lower Final Performance and Fewer Bits Flipped are Better.}
    \label{tab:comparison_all_methods}
    \begin{tabular}{@{}l|cc|cc|cc|cc@{}}
        \toprule
        & \multicolumn{2}{c|}{\textbf{FlipLLM (Proposed)}} & \multicolumn{2}{c|}{\color{blue}\textbf{GenBFA~\cite{das2024attentionbreaker}}} & \multicolumn{2}{c|}{\textbf{Gradient-Based}} & \multicolumn{2}{c}{\textbf{Random Flips}} \\
        \textbf{Model} & \textbf{Final Perf.} & \textbf{Crit. Bits} & {\color{blue}\textbf{Final Perf.}} & {\color{blue}\textbf{Crit. Bits}} & \textbf{Final Perf.} & \textbf{Crit. Bits} & \textbf{Final Perf.} & \textbf{Bits Flipped} \\
        \midrule
        LLaMA 3.1 8B     & \textbf{0.18\%} & \textbf{5} & {\color{blue}0.20\%} & {\color{blue}5} & 0.95\% & 850 & 67.5\% & 10,000 \\
        DeepSeek-V2      & \textbf{0.15\%} & \textbf{6} & {\color{blue}0.17\%} & {\color{blue}6} & 1.10\% & 920 & 69.0\% & 10,000 \\
        GPT-2 Large      & \textbf{0.35\%} & \textbf{5} & {\color{blue}0.38\%} & {\color{blue}5} & 1.50\% & 670 & 29.5\% & 10,000 \\
        LLaVA 1.6 (VLM)  & \textbf{0.5\%}  & \textbf{7} & {\color{blue}0.6\%} & {\color{blue}7} & 75.1\% & 1,150 & 77.8\% & 10,000 \\
        \bottomrule
    \end{tabular}
\end{table*}

\section{Results and Discussion}
\label{sec:results}
This section presents an extensive empirical evaluation and analysis of the proposed FlipLLM framework. We demonstrate its efficacy in identifying critical bit-flip vulnerabilities across diverse foundation models, quantify the resulting performance degradation, analyze the characteristics of the identified vulnerabilities, including their architectural localization, evaluate the framework's computational efficiency relative to baselines, and investigate the impact of key methodological choices through ablation studies.

\subsection{FlipLLM Effectiveness Across Diverse Foundation Models}
\label{subsec:results_main}
This section evaluates the core effectiveness of the FlipLLM framework by demonstrating its ability to induce significant functional degradation across various language and multimodal models using a minimal number of identified critical bit-flips.
Table~\ref{tab:main_results} summarizes the primary outcomes, showcasing FlipLLM's capability to achieve catastrophic performance collapse across different domains. In the language domain, targeting the widely used LLaMA 3.1 8B model, FlipLLM successfully reduces its MMLU accuracy from a baseline of \emph{69.9\%} to near-random chance (\emph{0.21\%}) using only \emph{5} bit-flips. The Mixture-of-Experts model, DeepSeek-V2, exhibits similar vulnerability, with its accuracy dropping from \emph{71.3\%} to \emph{0.19\%} after only \emph{6} flips. Even the smaller GPT-2 Large model was effectively neutralized, its accuracy falling from \emph{30.5\%} to \emph{0.35\%} with just \emph{5} flips. We observe similar behavior for all models on MMLU-Pro benchmark as well.
Crucially, we demonstrate that FlipLLM's effectiveness is not confined to text-only models. When applied to the LLaVA 1.6 multimodal model, FlipLLM identifies a similarly sparse set of critical bits. Flipping an average of only \emph{7} bits collapses LLaVA's performance on the VQAv2 benchmark from a robust \emph{78.2\%} to a non-functional \emph{0.5\%}. A similar degradation is observed on the more challenging TextVQA benchmark. In all instances, the small number of flipped bits represents an infinitesimal fraction (e.g., as low as \SI{6.23e-8}{\%} for LLaMA 3.1 8B) of the total bits comprising the model parameters, highlighting the extreme leverage and stealth of attacks identified efficiently by FlipLLM.

\subsection{Comparison with Baselines}
\label{subsec:results_comparison}

In this section, we contextualize FlipLLM's effectiveness by comparing its attack efficiency against both naive and classic BFA methodologies. This comparison, summarized in Table~\ref{tab:comparison_all_methods}, highlights the superiority of FlipLLM's adaptive, combinatorial search. Across both language and multimodal domains, FlipLLM consistently identifies hyper-efficient fault sets that catastrophically degrade model performance. For instance, using an average of only 5 critical bit-flips, FlipLLM drove LLaMA 3.1 8B's MMLU accuracy from 69.9\% down to 0.18\%. In contrast, the other methods proved far less potent. The gradient-based (DeepHammer-like) approach, while better than random, required a brute-force injection of 850 bits to achieve a comparable failure, demonstrating its inability to find the small, synergistic fault combinations discovered by FlipLLM. Even more starkly, applying 10,000 random flips within the same sensitive layer ($l^*$) barely impacted performance, resulting in an accuracy of 67.5\%. This trend holds for our multimodal target, LLaVA 1.6, where FlipLLM achieves near-zero VQA performance with only 7 bits, while random and gradient-based methods show negligible impact.
Figure~\ref{fig:FlipLLM_vs_random_combined} visually illustrates the cumulative impact as the identified critical bits are flipped one by one. For all three target models, MMLU accuracy remains relatively stable after one or two flips but suffers a catastrophic collapse to near-zero after a small threshold of 4-5 targeted MSB flips is reached. This demonstrates the extreme potency of the minimal bit sets identified by our framework.

\subsection{Computational Efficiency Comparison with SOTA}
\label{subsec:results_efficiency}

Beyond attack effectiveness, practical applicability hinges on computational efficiency. In this section, we systematically evaluate FlipLLM's runtime against the state-of-the-art evolutionary approach, GenBFA~\cite{das2024attentionbreaker}. This comparison, summarized in Table~\ref{tab:runtime_comparison}, highlights FlipLLM's significant speed advantage. We exclude the gradient-based baseline from this runtime analysis because, as shown in Table~\ref{tab:comparison_all_methods}, it requires hundreds of bits to degrade performance and fails to achieve the primary attack objective of catastrophic collapse with a minimal, stealthy fault set.
\emph{It is worth mentioning that GenBFA’s internal \emph{optimization} and \emph{sensitivity analysis}, primarily rely on a minimal proxy—performance on just the first batch with 4 samples of the `astronomy' MMLU task (approximately 1/1425\textsuperscript{th} of the MMLU test set). In contrast, FlipLLM uses a significantly larger subset with 570 samples (1/10\textsuperscript{th} of the MMLU test set) for reward calculations and sensitivity profiling prior to full-suite evaluation. This broader evaluation scope contributes to the discovery of more generalizable vulnerabilities.}
To facilitate a fair runtime comparison, we implemented GenBFA in our computational environment using the parameters reported in their work, and measured the time each method required to identify a minimal critical bit set of the same size that induces catastrophic failure.

FlipLLM exhibits markedly improved computational performance and completes its end-to-end search in \textbf{18 hours} for LLaMA 3.1 8B, \textbf{26 hours} for DeepSeek-V2, and \textbf{4 hours} for GPT-2 Large. The efficiency advantage extends to the multimodal domain, with FlipLLM requiring only \textbf{22 hours} to analyze LLaVA 1.6. In contrast, GenBFA requires a protracted \textbf{43 hours} for LLaMA 3.1, \textbf{42 hours} for DeepSeek-V2, \textbf{10 hours} for GPT-2 Large, and \textbf{48 hours} for LLaVA 1.6. These results confirm that FlipLLM delivers superior attack effectiveness with significantly reduced computational cost—achieving up to \textbf{2.5$\times$ faster runtime} on GPT-2 Large and an average of \textbf{1.9$\times$ faster runtime} across all evaluated models.
This efficiency stems from FlipLLM's core design. Its Q-learning agent incrementally refines a bit-selection policy based on sequential model interactions, enabling more targeted and sample-efficient exploration. In contrast, GenBFA's reliance on population-based evolutionary search requires evaluating many candidate solutions per generation, which scales less effectively. FlipLLM’s computational gains are further enabled by its two-stage design: initial sensitivity profiling aggressively prunes the search space, allowing the subsequent reward-driven RL to explore this reduced space with high efficiency. These components make FlipLLM a significantly more scalable and practical framework for conducting vulnerability analysis on large-scale foundation models.

\begin{table}[!t]
    \centering
    \caption{End-to-End Runtime (Hours) for FlipLLM vs. GenBFA.}
    \label{tab:runtime_comparison}
    \resizebox{\columnwidth}{!}{%
    \begin{tabular}{@{}lcccc@{}}
        \toprule
        \textbf{Method} & \textbf{LLaMA 3.1} & \textbf{DeepSeek-V2} & \textbf{GPT-2 L} & \textbf{LLaVA 1.6} \\
        \cmidrule(lr){2-2} \cmidrule(lr){3-3} \cmidrule(lr){4-4} \cmidrule(lr){5-5}
        & (8B) & (7B MoE) & (774M) & (7B VLM) \\
        \midrule
        \textbf{FlipLLM (Ours)} & \textbf{18} & \textbf{26} & \textbf{4} & \textbf{22} \\
        GenBFA~\cite{das2024attentionbreaker} & 43 & 42 & 10 & 48 \\
        \bottomrule
    \end{tabular}
    }
\end{table}

\subsection{Framework Scalability Analysis}
\label{subsec:consistency_scalability}
\begin{figure}[t]
    \centering
    \includegraphics[width=0.9\columnwidth]{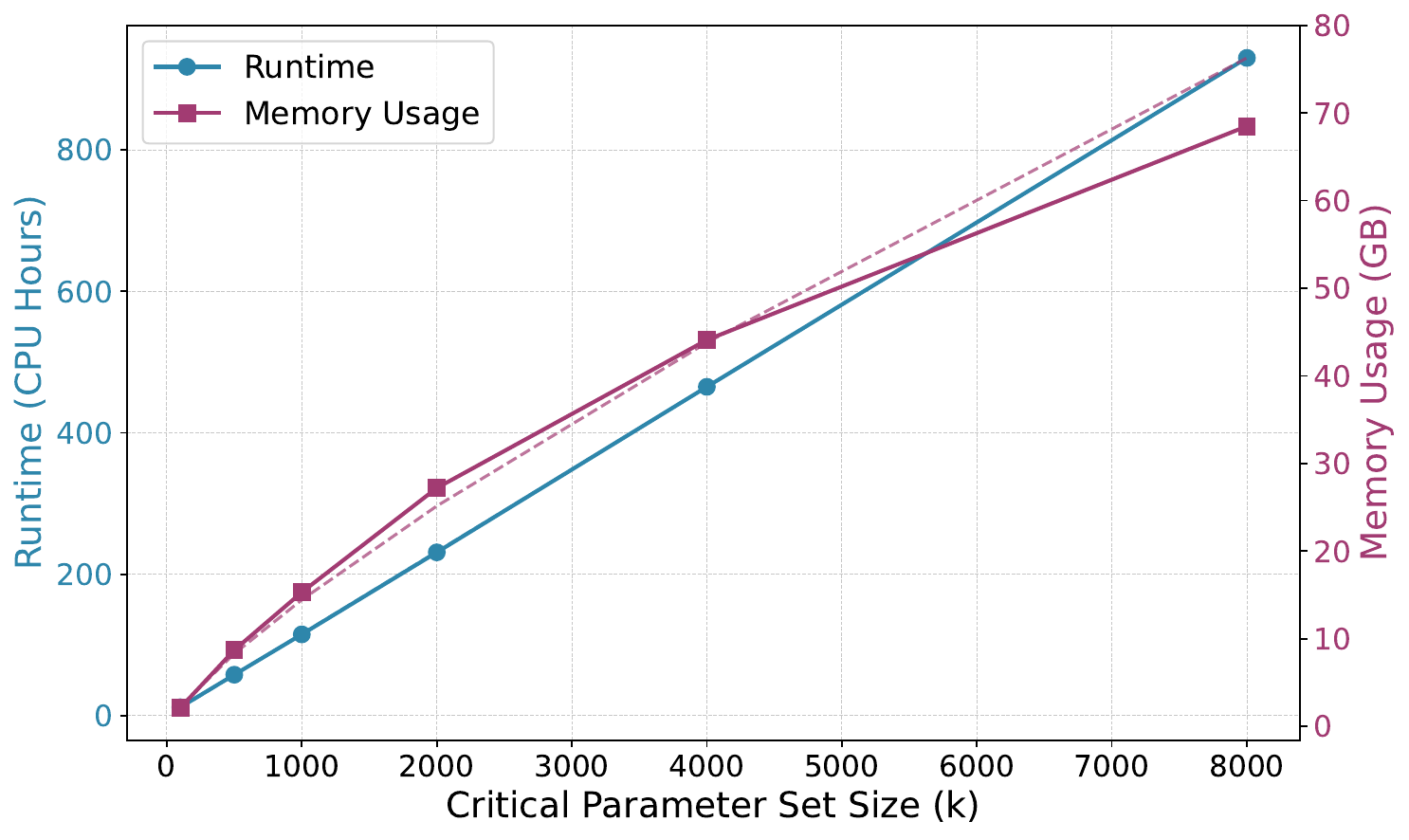} 
    \caption{Computational scalability of FlipLLM's Q-Learning phase on LLaMA 3.1 8B, showing linear runtime growth ($R^2 > 0.99$) and predictable, super-linear memory usage ($O(k^{1.3})$) with respect to the initial candidate parameter set size, $k$.
    }
    \vspace{-2mm}
    \label{fig:scalability_results}
\end{figure}

Beyond raw speed, a practical BFA discovery framework must be scalable, meaning its computational cost should not grow prohibitively with the complexity of the target model or the size of the search space. We analyzed FlipLLM's computational scaling characteristics by varying the size of the initial candidate parameter set ($k = |I_{\text{subset}}^{l^*}|$) that the RL agent explores during Phase 3. The results for LLaMA 3.1 8B, presented in Figure~\ref{fig:scalability_results}, confirm the predictable and efficient scaling of our approach.

The attack runtime exhibits a strong \textbf{linear relationship} with the number of candidate parameters $k$, as evidenced by a near-perfect linear fit ($R^2 > 0.99$). This is a critical property, demonstrating that doubling the search space for the RL agent only doubles the runtime, avoiding the exponential scaling that plagues exhaustive search methods. The framework's memory requirements grow at a modest, \textbf{super-linear rate}, empirically determined to be approximately $O(k^{1.3})$. This predictable, non-exponential scaling is a direct result of FlipLLM's two-phase design, where the expensive RL exploration is constrained to a pre-qualified subset of parameters. To provide a concrete data point, discovering the minimal critical fault set $|I_{\text{critical}}|$ from an initial candidate set of $k = 8{,}000$ sensitive parameters in LLaMA 3.1 8B was completed in approximately 18 GPU hours. This demonstrates that FlipLLM is a computationally feasible methodology for discovering vulnerabilities in state-of-the-art, billion-parameter foundation models.

\subsection{Analysis of Internal FlipLLM Phases}
\label{subsec:results_phases}

In this section, we analyze the internal behavior of the FlipLLM framework to evaluate the contributions of its individual phases: sensitivity-guided layer selection (Phase 1) and RL optimization (Phase 3). This analysis helps to explain how FlipLLM achieves efficient and minimal bit-flip identification. While the detailed analysis is presented for LLaMA 3.1 8B, we observed similar internal dynamics across other evaluated models, including DeepSeek-V2, GPT-2 Large, and the multimodal LLaVA 1.6.

\subsubsection{Sensitivity Profiling and Layer Selection}
\begin{figure}[!t]
    \centering
    \includegraphics[width=0.95\columnwidth]{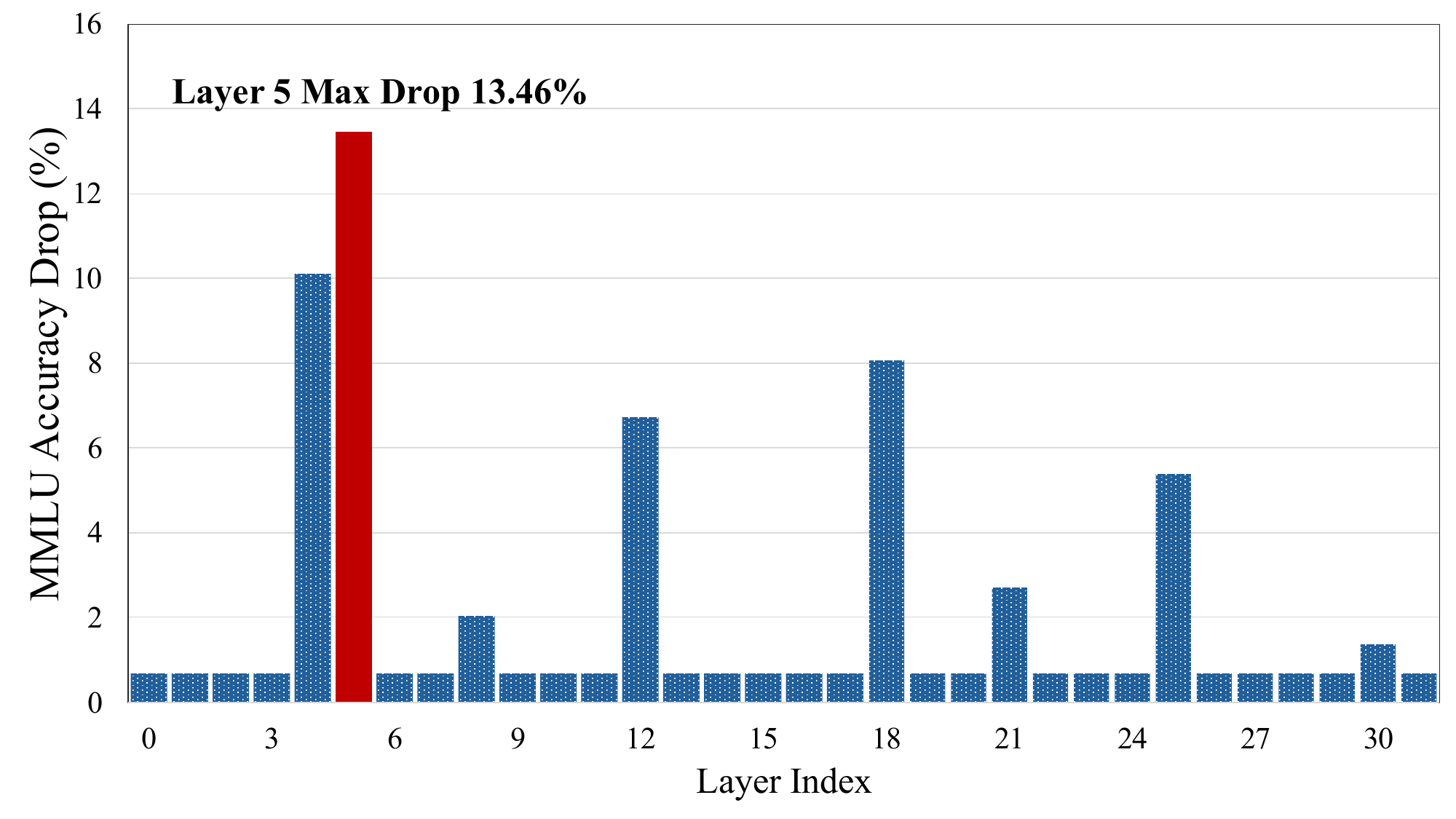} 
    \caption{Layer sensitivity profiling for LLaMA 3.1 8B, showing accuracy after perturbing the top 0.1\% sensitive parameters in each layer. Attention and normalization layers show the highest vulnerability.}
    \label{fig:sensitivity_results_detail}
\end{figure}

The layer-wise sensitivity analysis (Algorithm~\ref{alg:sensitivity_profiling}) effectively localizes vulnerability to a single layer, drastically pruning the search space. Figure~\ref{fig:sensitivity_results_detail} presents representative results for LLaMA 3.1 8B, showing the initial accuracy drop caused by perturbing the top 0.1\% sensitive parameters (identified via Eq.~\eqref{eq:sensitivity}) in different layers. A distinct pattern emerges: layers containing normalization and attention projection weights exhibited the most significant accuracy degradation upon perturbation, whereas layers dominated by feed-forward network (FFN) parameters showed relatively minor impact. The layers ($l^*$) selected based on the maximum initial accuracy drop consistently proved to be a high-potential target for the subsequent RL optimization, validating this crucial first step of our framework.

\subsubsection{RL Optimization Dynamics}
\begin{figure}[!b]
    \centering
    \includegraphics[width=0.95\columnwidth]{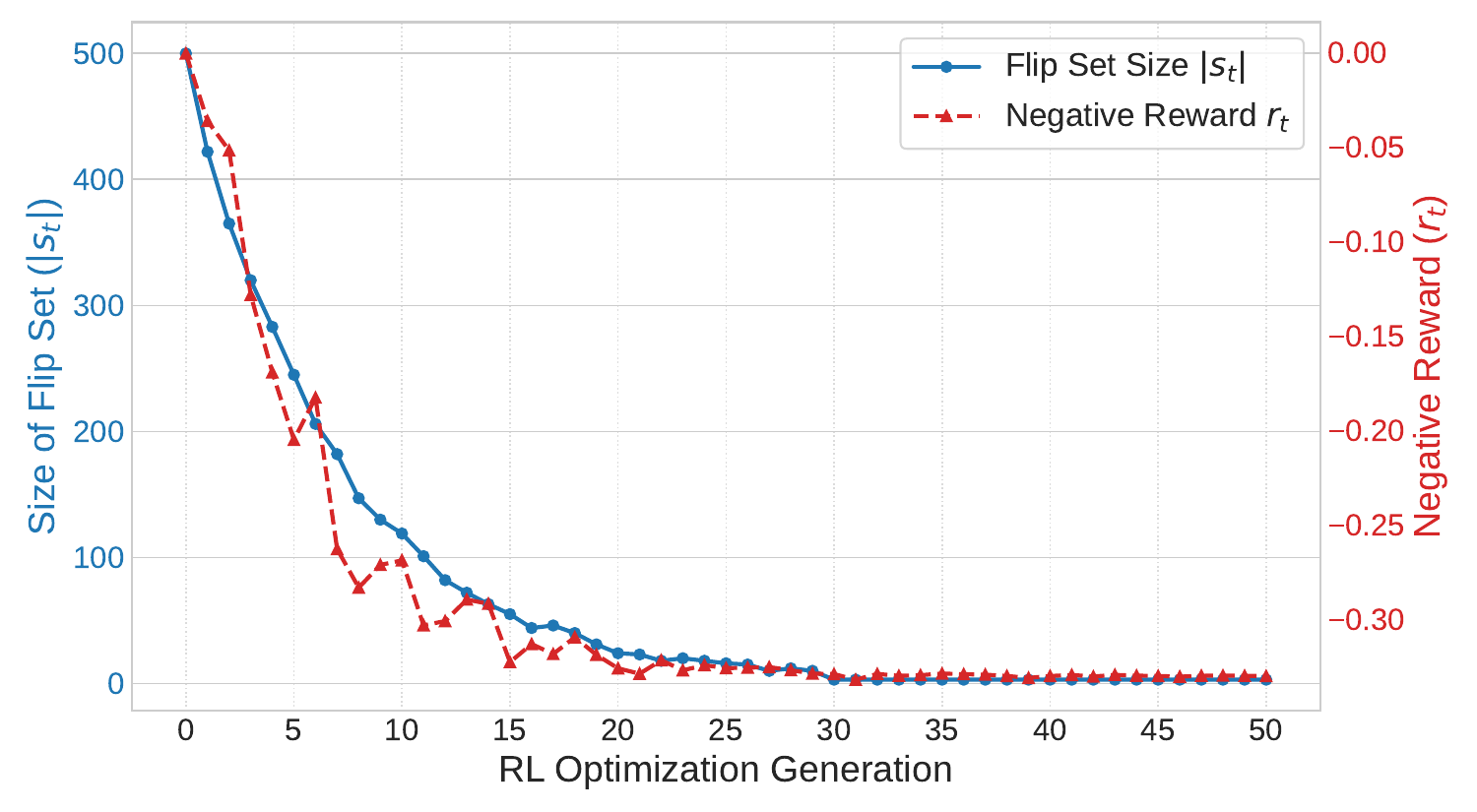} 
    \caption{FlipLLM Q-learning dynamics on LLaMA 3.1 8B over 50 generations. The agent rapidly reduces the critical bit set size ($|s_t|$, purple) while improving the impact-per-flip (approximated by negative reward, red).}
    \label{fig:rl_dynamics}
\end{figure}

The Q-learning optimization phase refines the initial sensitive subset ($I_{\text{hybrid}}^{l^*}$) into the minimal critical set ($I_{\text{critical}}$). Figure~\ref{fig:rl_dynamics} illustrates the agent's learning trajectory when attacking LLaMA 3.1 8B. It plots the negative reward (approximating impact per flip, $ (1-\emph{acc}_t)/|s_t| $) and the size of the active flip set ($|s_t|$) over the configured budget of $G = 50$ RL generations. The plot shows that the critical flip set size $|s_t|$ (purple, left y-axis) rapidly decreases and stabilizes near the final minimal set of size 5. Concurrently, the negative reward (red, right y-axis) trends downward, indicating the agent is successfully finding smaller bit sets that cause greater performance degradation per flip. This behavior reflects a successful policy adaptation toward sparse, high-impact perturbations. The convergence on a small, highly effective flip set $I_{\text{critical}}$ validates the formulation of the RL phase and the efficacy of the reward function (Eq.~\eqref{eq:reward}) in driving minimal, high-leverage attacks.

\textcolor{blue}{\subsection{Evaluation Against Hardware Protection Mechanisms}}
\label{subsec:defense_eval}

\textcolor{blue}{In this subsection we access the effectiveness of our framework in realistic deployments scenarios. 
Table~\ref{tab:defense_effectiveness} shows the FlipLLM efficiancy against the Unprotected hardware and Single Error Correction, Double Error Detection (ECC SECDED) protected hardware reliability mechanisms. Since FlipLLM's minimal critical fault sets consist of a small number of single-bit flips spread across different memory words, a standard ECC implementation can correct each flip independently. Our analysis confirms the following. For LLaMA 3.1 8B, an ECC-protected system maintains \textbf{69.8\% accuracy} (compared to a 69.9\% baseline), completely neutralizing the attack that reduces an unprotected system to 0.21\% accuracy. This demonstrates that implementing ECC in FlipLLM identified locations is a highly effective countermeasure against such targeted attacks.}
\textcolor{blue}{This analysis provides two critical insights. First, it underscores the importance of deploying hardware-level protections in any security or safety-critical AI system. Our work quantifies the catastrophic risk of omitting such defenses in cost-sensitive edge devices. Second, it demonstrates the dual-use nature of the FlipLLM framework. For an attacker, it identifies the most potent vulnerabilities to target. For a \textit{hardware designer (defender)}, it provides a powerful tool to perform a worst-case analysis, justifying the inclusion of defenses like ECC and guiding the development of more cost-effective, selective protection schemes that can achieve high coverage with lower overhead.}

\begin{table}[!t]
    \centering
    \caption{\textcolor{blue}{Effectiveness of Hardware Protections Against FlipLLM}}
    \label{tab:defense_effectiveness}
    \resizebox{\columnwidth}{!}{%
    \begin{tabular}{@{}lccc@{}}
        \toprule
        \textbf{System Configuration} & \textbf{LLaMA 3.1 8B} & \textbf{DeepSeek-V2} & \textbf{GPT-2 Large} \\
        \textbf{} & \textbf{(MMLU Acc.)} & \textbf{(MMLU Acc.)} & \textbf{(MMLU Acc.)} \\
        \midrule
        \textcolor{blue}{Baseline (Original)} & \textcolor{blue}{69.9\%} & \textcolor{blue}{71.3\%} & \textcolor{blue}{30.5\%} \\
        \textcolor{blue}{Unprotected + FlipLLM} & \textcolor{blue}{0.21\%} & \textcolor{blue}{0.19\%} & \textcolor{blue}{0.35\%} \\
        \textcolor{blue}{\textbf{\makecell[l]{ECC SECDED \\Protected + FlipLLM}}} & \textcolor{blue}{\textbf{69.8\%}} & \textcolor{blue}{\textbf{71.2\%}} & \textcolor{blue}{\textbf{30.4\%}} \\
        \bottomrule
    \end{tabular}
    }
\end{table}

\vspace{2mm}
\subsection{Ablation Studies and Parameter Sensitivity}
\label{subsec:results_ablation}

To assess FlipLLM's robustness and the impact of its design choices, we performed ablation studies focusing on key components and hyperparameters. Due to computational cost constraints, these studies were primarily conducted on LLaMA 3.1 8B, though we believe this trend can be generalized to other LLMs as well.

\subsubsection{Impact of Sensitivity Metric (\(\alpha\))}
\begin{figure}[!t]
    \centering
    \includegraphics[width=0.9\linewidth]{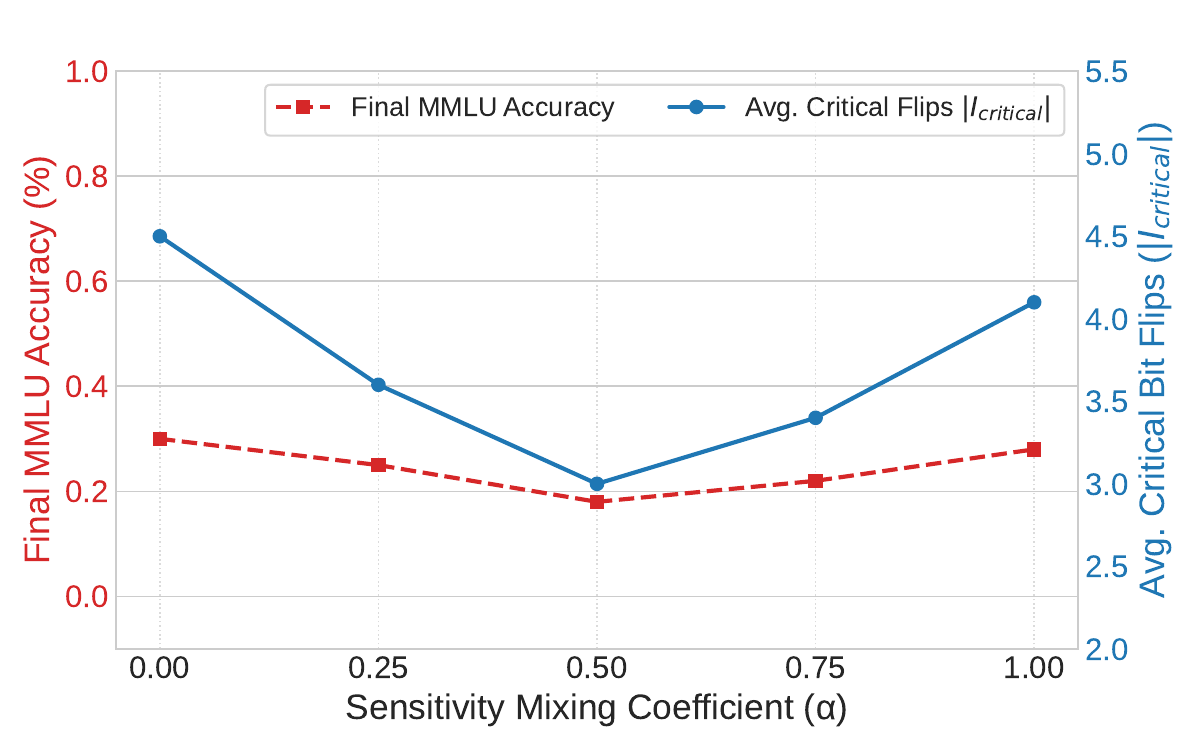}
    \caption{Impact of Sensitivity Mixing Coefficient \(\alpha\) on FlipLLM for LLaMA 3.1 8B. 
    }
    \label{fig:ablation_alpha} 
    \vspace{-2mm}
\end{figure}

We evaluated FlipLLM's final performance while varying the sensitivity mixing coefficient \(\alpha\) (Eq.~\eqref{eq:sensitivity}) from 0 (magnitude-only) to 1 (gradient-only). Figure~\ref{fig:ablation_alpha} shows the resulting final MMLU accuracy and the average size of \(|I_{\text{critical}}|\). While pure magnitude (\(\alpha=0\)) and pure gradient (\(\alpha=1\)) approaches could still identify damaging flips, the hybrid approach (\(\alpha=0.5\)) consistently found the smallest critical sets (\(|I_{\text{critical}}|\) averaging \textbf{5.0}) for achieving close to zero accuracy. Using only magnitude (\(\alpha=0\)) required slightly more flips (avg. \textbf{7.1}), while using only gradient (\(\alpha=1\)) required avg. \textbf{6.4} flips. This validates the benefit of combining both static (magnitude) and dynamic (gradient) information for initialization. Performance degradation was severe across all tested \(\alpha\) values, but the hybrid metric yielded the most efficient attack set.

\subsubsection{Impact of Guided Search (Phases 1 \& 2)}
We compare the standard FlipLLM pipeline against an \emph{RL-Only} baseline to quantify the benefit of the initial sensitivity profiling and subset selection. In this baseline, the Q-learning agent (Phase 3) started its search from a \emph{randomly} selected initial subset of the same size (\(k\)) within the same sensitive layer \(l^*\), effectively skipping the guidance from Phases 1 and 2. The RL-Only baseline failed to converge to a comparably small critical set within the standard \(G=50\) generations. Achieving a similar level of accuracy collapse required either significantly more generations (\(G \approx 1000\), estimated) or resulted in a much larger critical set (\(|I_{\text{critical}}| \approx 500\), estimated) after 50 generations. This highlights the crucial role of the sensitivity-guided initialization (Phases 1 and 2) in accelerating convergence and enabling the discovery of highly parsimonious solutions within a practical computational budget.

\subsubsection{Impact of RL Hyperparameters}
We investigated sensitivity to the number of RL generations (\(G\)) and exploration rate (\(\epsilon\)). Increasing \(G\) beyond 50 produced diminishing returns, reducing \(|I_{\text{critical}}|\) by less than 0.3 flips on average. Reducing \(G\) to 10 led to noticeably larger critical sets (avg. 60.2 flips). Varying \(\epsilon\) between 0.05 and 0.2 did not significantly alter the final average \(|I_{\text{critical}}|\), suggesting reasonable stability. The chosen \(G=50\) and \(\epsilon=0.1\) offer a good balance between solution quality and computational cost.

These ablation studies validate the importance of FlipLLM’s core design elements. The hybrid sensitivity score improves initialization efficiency. Sensitivity-guided search significantly reduces the number of required RL iterations. The Q-learning phase is effective and stable under moderate hyperparameter variation, confirming the overall robustness and scalability of the FlipLLM framework.

\subsubsection{Reinforcement Learning Algorithm Analysis}
\label{subsec:rl_ablation}

\textcolor{blue}{To validate our choice of Q-learning, we performed an ablation study comparing its performance against two simpler, non-learning heuristics on the LLaMA 3.1 8B model. The results, summarized in Table~\ref{tab:rl_comparison}, demonstrate the critical importance of a learning-based approach for this problem. A purely \emph{Random Search} approach is highly inefficient, requiring 47 bits to cause a partial accuracy drop. A \emph{Greedy Selection} strategy, which always chooses the action with the highest immediate reward, performs better but converges to a local optimum, requiring 15 bits. In contrast, \textit{Q-learning (FlipLLM)} discovers a far more potent and minimal fault set of only \emph{5 bits}.}
\textcolor{blue}{This superior performance stems from Q-learning's ability to balance exploration and exploitation, allowing it to discover non-obvious, synergistic fault combinations that myopic greedy approaches miss. While more advanced RL algorithms such as Policy Gradient or Actor-Critic methods exist, we chose Q-learning for two primary reasons. First, our problem is defined by a discrete action space ($\{\text{add, remove, shift}\}$) and, after our Phase 1-2 pruning, a tractable state space, a domain where tabular Q-learning is highly sample-efficient and guaranteed to converge. Second, it provides an excellent balance of high performance and implementation simplicity.
}

\begin{table}[!t]
    \centering
    \caption{\textcolor{blue}{RL Algorithm Ablation Study on LLaMA 3.1 8B.}}
    \label{tab:rl_comparison}
    \resizebox{\columnwidth}{!}{%
    \begin{tabular}{@{}lccc@{}}
        \toprule
        \textbf{Methods} & \textbf{\makecell[c]{Final \\Accuracy (\%)}} & \textbf{\makecell[c]{Critical Bits\\ Found}} & \textbf{\makecell[c]{Runtime\\ (hrs)}} \\
        \midrule
        \textcolor{blue}{Random Search} & \textcolor{blue}{12.3\%} & \textcolor{blue}{47} & \textcolor{blue}{8.5} \\
        \textcolor{blue}{Greedy Selection} & \textcolor{blue}{2.8\%} & \textcolor{blue}{15} & \textcolor{blue}{12.0} \\
        \textcolor{blue}{\textbf{Q-Learning (FlipLLM)}} & \textcolor{blue}{\textbf{0.21\%}} & \textcolor{blue}{\textbf{5}} & \textcolor{blue}{18.0} \\
        \bottomrule
    \end{tabular}
    }
    \vspace{-1mm}
\end{table}

\subsection{Architectural Vulnerability Localization \& Hardware Implications}
\label{subsec:results_localization}
Across all tested models, FlipLLM consistently identified critical bit-flip vulnerabilities localized within specific architectural components—primarily Layer Normalization (or RMS Norm) layers and Attention projection matrices (Query, Key, Value, and Output). Feed-forward network (FFN) parameters were rarely selected, suggesting a markedly uneven vulnerability distribution across model submodules. This observation enables fine-grained hardware introspection: rather than applying uniform protection, system architects can prioritize resilience measures for memory segments and compute units responsible for normalization and attention projections. Selective enhancements such as ECC, parity, or redundancy (e.g., TMR) could be focused on these areas to reduce hardware overhead without compromising security or reliability.

Furthermore, the exact indices identified by FlipLLM offer realistic, empirically grounded fault models for CAD verification workflows. These minimal, high-impact flip sets can drive targeted fault injection campaigns using industry-standard environments (e.g., UVM, SystemVerilog, Synopsys Z01X). Such campaigns enable rigorous pre-silicon validation against worst-case bit-flip scenarios, substantially improving upon traditional Single Event Upset (SEU) models. FlipLLM also supports proactive vulnerability assessment during early-stage hardware design by analyzing candidate architectures in simulation or emulation environments. The resulting \(|I_{\text{critical}}|\) metric provides a quantifiable measure of fault tolerance, facilitating comparison across design choices, quantization strategies, or protection schemes. In doing so, FlipLLM establishes a principled bridge between high-level algorithmic fault analysis and hardware-level design decisions.

Future work will explore extending FlipLLM to multi-layer or cross-layer fault localization. We also aim to integrate FlipLLM within hardware-in-the-loop simulation environments to co-optimize LLM robustness and silicon-level fault tolerance.
\textcolor{blue}{\subsection{Scalability Analysis for Large Scale Models}}
\label{subsec:scalability}

\textcolor{blue}{FlipLLM's computational requirements follow predictable scaling patterns that enable deployment across diverse model sizes. This section presents empirical scaling characteristics, theoretical projections for larger models, limitations and potential future directions.}

\textcolor{blue}{\subsubsection{Computational Complexity Analysis.} The total computational cost of FlipLLM comprises two components: (1) model evaluation overhead for forward/backward passes during sensitivity profiling and RL optimization, and (2) Q-learning exploration over the pruned search space of size $k$. Empirical measurements across models ranging from 774M to 8B parameters reveal that component (2) dominates, exhibiting strong linear scaling with search space size ($R^2 > 0.99$). Component (1) scales with model parameter count $N$, but modern inference optimizations~\cite{wan2024efficientlargelanguagemodels} such as grouped-query attention, KV-cache compression, and quantization demonstrate sub-linear scaling behavior in practice, particularly for models above 10B parameters where these optimizations become standard.}
\textcolor{blue}{The search space size $k$ is determined by the selection rate $r$ (set to 0.1\% in our experiments): $k = r \cdot N$. For a model with $N$ parameters, the expected runtime scales approximately as:}
\begin{equation}
\small
\textcolor{blue}{T(N) \approx \xi \cdot N^{\eta} + \lambda \cdot k \cdot G}
\end{equation}
\textcolor{blue}{where $\xi$ and $\lambda$ are constants derived from hardware characteristics, $\eta \in [0.7, 1.0]$ reflects sub-linear inference scaling, $G$ is the number of RL episodes, and the second term dominates for $N > 1B$.}

\textcolor{blue}{\subsubsection{Scaling Analysis Across Model Sizes.} Table~\ref{tab:scaling_projections} presents FlipLLM's runtime requirements across model sizes ranging from 774M to 120B parameters. Empirical measurements on GPT-2 Large (774M), DeepSeek-V2 7B, and LLaMA 3.1 8B demonstrate runtimes of 4, 26, and 18 GPU hours respectively. For larger models, conservative projections based on observed linear scaling with search space size $k$ estimate 2,700--3,400 GPU hours for Gemma 2 27B, 8,000--10,000 GPU hours for LLaMA 3.1 70B, and 14,000--17,000 GPU hours for GPT-4 class systems (120B parameters). These projections assume maintenance of the 0.1\% selection rate and linear scaling of the RL optimization component, both validated empirically across the 774M--8B range. Notably, FlipLLM's requirements remain orders of magnitude below exhaustive search (search space $> 10^{48}$) while achieving 2.5× faster runtime than state-of-the-art evolutionary methods.}


\begin{table}[h]
\centering
\caption{\textcolor{blue}{FlipLLM Scaling Analysis: Runtime and Infrastructure Requirements}}
\label{tab:scaling_projections}
\resizebox{\columnwidth}{!}{
\begin{tabular}{lccc}
\toprule
\textbf{\textcolor{blue}{Model}} & \textbf{\textcolor{blue}{Parameters}} & \textbf{\textcolor{blue}{Search Space}} & \textbf{\textcolor{blue}{GPU Hours}} \\
 & \textbf{\textcolor{blue}{($N$)}} & \textbf{\textcolor{blue}{($k$)}} &  \\
\midrule
\textcolor{blue}{GPT-2 Large} & \textcolor{blue}{774M} & \textcolor{blue}{774} & \textcolor{blue}{4} \\
\textcolor{blue}{DeepSeek-V2 7B} & \textcolor{blue}{7B} & \textcolor{blue}{7,000} & \textcolor{blue}{26} \\
\textcolor{blue}{LLaMA 3.1 8B} & \textcolor{blue}{8B} & \textcolor{blue}{8,000} & \textcolor{blue}{18} \\
\textcolor{blue}{Gemma 2 27B} & \textcolor{blue}{27B} & \textcolor{blue}{27,000} & \textcolor{blue}{2,700--3,400} \\
\textcolor{blue}{LLaMA 3.1 70B} & \textcolor{blue}{70B} & \textcolor{blue}{70,000} & \textcolor{blue}{8,000--10,000} \\
\textcolor{blue}{GPT-4 Class (Est.)} & \textcolor{blue}{120B} & \textcolor{blue}{120,000} & \textcolor{blue}{14,000--17,000} \\
\bottomrule
\end{tabular}
}
\end{table}

\textcolor{blue}{\subsubsection{Architectural Consistency Across Scales.} Analysis across the 10× parameter range from GPT-2 Large (774M) to LLaMA 3.1 8B reveals consistent localization of critical bits in attention projection matrices (Query, Key, Value, Output) and normalization layers (LayerNorm, RMSNorm).
This pattern holds across dense transformers (GPT-2, LLaMA 3.1), mixture-of-experts models (DeepSeek-V2), and multimodal architectures (LLaVA-1.5), indicating that vulnerability patterns are architecture-agnostic rather than scale-dependent. The Q-learning optimization successfully converges to minimal critical sets within 50 episodes for search spaces up to $k = 8{,}000$ parameters. The two-phase design maintains scalability: Phase 1 sensitivity profiling reduces the search space by 99.9\% before RL optimization, ensuring manageable action spaces regardless of model size.}

\vspace{2mm}
\section{Conclusion}                
\label{sec:conclusion}

This paper introduced \textbf{FlipLLM}, a reinforcement learning-based framework designed to address the scalability and generalizability challenges of discovering bit-flip attack (BFA) vulnerabilities in modern foundation models. By formulating BFA discovery as a sequential decision-making problem, FlipLLM efficiently identifies minimal, high-impact bit sets capable of inducing catastrophic failure through a combination of sensitivity-guided layer pruning and Q-learning.
Our experiments demonstrated the effectiveness and generalizability of FlipLLM across a diverse set of models. We evaluated prominent text-only LLMs including GPT-2 Large, LLaMA 3.1 8B, and DeepSeek-V2 7B, as well as Large Vision Models (VLMs) like LLaVA 1.6. The results confirm that FlipLLM is a potent attack discovery tool: on the MMLU and MMLU-Pro benchmarks, flipping as few as 5 bits identified by our method plummeted the accuracy of LLaMA 3.1 8B from 69.9\% to $\approx$ 0.2\%. We showed this vulnerability extends to the multimodal domain, where FlipLLM collapsed LLaVA's VQA score on VQAv2 and TextVQA from over 78\% to almost 0\% with a similar 7 number of flips. Furthermore, our framework provides these results with high efficiency, identifying critical bits up to \textbf{2.5$\times$} faster than state-of-the-art methods. In conclusion, FlipLLM offers the first scalable and adaptive methodology for exploring the BFA vulnerability of both language and multimodal foundation models. This work not only provides a powerful new tool for security researchers but also paves the way for more comprehensive hardware-security evaluation, a critical step towards building robust and reliable AI systems.
\textcolor{blue}{In the future work, we will explore more advanced RL methods (e.g., policy gradient, actor-critic) or meta-learning approaches that leverage cross-model patterns to further improve scalability.}

\section{ACKNOWLEDGMENTS}
This material is based upon work supported by the National Science Foundation (NSF) under Award Numbers: CCF-2323819. Any opinions, findings, conclusions, or recommendations expressed in this publication are those of the authors and do not necessarily reflect the views of the NSF.

\bibliographystyle{IEEEtran}
\bibliography{bib/conf}

\end{document}